# Unidirectional Kondo scattering in layered NbS$_2$

Edoardo Martino*[1], Carsten Putzke[2], Markus König[3], Philip Moll[2], Helmuth Berger[1], David LeBoeuf[4], Maxime Leroux[4], Cyril Proust[4], Ana Akrap[5], Holm Kirmse[6], Christoph Koch[6], ShengNan Zhang[1,7], QuanSheng Wu[1,7], Oleg V. Yazyev[1,7], László Forró[1,8], Konstantin Semeniuk*[1]

1) *Institute of Physics, École Polytechnique Fédérale de Lausanne (EPFL), CH-1015 Lausanne, Switzerland*

2) *Institute of Materials Science and Engineering, École Polytechnique Fédérale de Lausanne (EPFL), CH-1015 Lausanne, Switzerland*

3) *Max Planck Institute for Chemical Physics of Solids, 01187 Dresden, Germany*

4) *Laboratoire National des Champs Magnétiques Intenses (LNCMI-EMFL), CNRS, UGA, UPS, INSA, Grenoble/Toulouse, France*

5) *University of Fribourg, Department of Physics, CH-1700 Fribourg, Switzerland*

6) *Humboldt University of Berlin, Department of Physics, Berlin 12489, Germany*

7) *National Center for Computational Design and Discovery of Novel Materials MARVEL, École Polytechnique Fédérale de Lausanne (EPFL), CH-1015 Lausanne, Switzerland*

8) *Stavropoulos Center for Complex Quantum Matter, University of Notre Dame, Notre Dame 46556 IN, USA*

*\* Emails for correspondence: edoardo.martino91@gmail.com, konstantinms@gmail.com*

**Abstract**

**Crystalline defects can modify quantum interactions in solids, causing unintuitive, even favourable, properties such as quantum Hall effect or superconducting vortex pinning. Here we present another example of this notion — an unexpected unidirectional Kondo scattering in single crystals of 2H-NbS$_2$. This manifests as a pronounced low-temperature enhancement in the out-of-plane resistivity and thermopower below 40 K, hidden for the in-plane charge transport. The anomaly can be suppressed by the c-axis-oriented magnetic field, but is unaffected by field applied along the planes. The magnetic moments originate from**

**layers of 1T-NbS$_2$, which inevitably form during the growth, undergoing a charge-density-wave reconstruction with each superlattice cell (David-star-shaped cluster of Nb atoms) hosting a localised spin. Our results demonstrate the unique and highly anisotropic response of a spontaneously formed Kondo-lattice heterostructure, intercalated in a layered conductor.**

## Introduction

Layered van der Waals materials, such as transition metal dichalcogenides (TMDs), have attracted major interest thanks to their rich variety of ground states and the possibility of their exfoliation down to an atomically thin level, which remarkably modifies their electronic properties[1,2]. Recent observations of intriguing physics in artificially-assembled heterostructures highlight the importance of interlayer interactions. Examples include the outstanding stability of interlayer excitons in semiconducting TMDs[3], and strongly correlated states in twisted bilayer systems[4]. Relevant aspects of the inter-plane coupling can be deduced by probing out-of-plane charge transport, even in bulk materials[5]. However, enforcing the current flow strictly along the *c* axis can be rather challenging due to the crystals' common flake-like appearance and their propensity for delamination. Such a pitfall can distort measurement results by orders of magnitude, as demonstrated in our recent study of microstructured samples of 1T-TaS$_2$ with a well-defined current flow[6]. This observation motivates a careful re-examination of the out-of-plane charge transport properties in this class of materials by adopting the latest state-of-the-art for quantum matter microfabrication[7].

Here we present data on the out-of-plane electrical resistivity of bulk monocrystalline 2H-NbS$_2$. This material is one of the three known structural variants of layered NbS$_2$. The two other polytypes are 3R and 1T, the latter occurring only in atomically thin form[8,9]. As illustrated in Figure 1, 1T-NbS$_2$ consists of corner-sharing octahedral NbS$_6$ cells. Layers of 2H- and 3R-NbS$_2$ both contain NbS$_6$ units of trigonal prismatic geometry, but exhibit different stacking configurations. The 1T polytype has been attracting interest recently as a candidate for realising a two-dimensional magnetic system[11,12]. 2H-NbS$_2$ has been actively featured in the literature due to a superconductivity below 6 K, proposed to have a multiband character. It also does not show any charge density wave (CDW) order, which is uncommon for metallic TMDs[13–18]. Another distinguishing feature of 2H-NbS$_2$ is its non-trivial synthesis procedure. This polytype is thermodynamically stable in a relatively narrow range of temperatures and reactant stoichiometries[19–21]. Crystals formed during high-temperature growth must be rapidly quenched in order to capture 2H-NbS$_2$ in a metastable room-temperature state. However, X-ray diffraction studies have shown that the resultant material has up to 18% of pairs of neighbouring layers stacked in a 3R-like manner[22,23]. Additionally, diffuse X-ray scattering experiments[23] revealed weak traces of the $\sqrt{13} \times \sqrt{13}$ CDW reconstruction, which appears as a triangular superlattice of David-star-shaped clusters defined by 13 Nb atoms[24]. Such a reconstruction is not expected for pure 2H-NbS$_2$ or 3R-NbS$_2$. Earlier theoretical investigations have predicted 1T-NbS$_2$ to be particularly prone to developing such a CDW order[11,12]. One can therefore conclude that single crystals of 2H-NbS$_2$ contain rare, atomically thin inclusions of the 1T polytype.

Our study of 2H-NbS$_2$ revealed a remarkably strong low-temperature anomaly in the compound's out-of-plane resistivity ($\rho_c$), manifesting as a minimum at around 40 K, followed by a pronounced upturn upon further cooling. The feature is simultaneously invisible in the in-plane resistivity ($\rho_{ab}$), and shows a highly anisotropic response to magnetic field. Neither 2H-NbSe$_2$ nor 3R-NbS$_2$ display such an anomaly, implying that the phenomenon is linked to the structural defects specific to 2H-NbS$_2$. 1T-NbS$_2$, layers of which are one of such defects, were predicted to form a lattice of unpaired localised spins located at the centre of each David-star CDW superlattice cluster[11,12]. We argue that planes of magnetic moments, hosted by the inclusions of 1T-NbS$_2$, cause a Kondo effect observable only when the current flows across these planes.

**Results**

Optimisation of sample geometry with focused ion beam (FIB) micro-milling greatly improves charge transport study precision[6,7]. Using this approach, we shaped single crystals into samples with well-defined few micron thick and wide current channels, oriented along the two principal directions: normal and parallel to the atomic planes (Figure 2a shows a sample of 2H-NbS$_2$ produced this way). Such a design allowed simultaneous measurement of both $\rho_{ab}$ and $\rho_c$ via the four-point technique. Probing $\rho_c$ on two segments of different surface-to-volume resulted in mutually consistent values, allowing us to ensure that our results were not distorted by the presence of surface-related effects or macroscopic defects.

Figure 2b shows the plots of $\rho_{ab}$ and $\rho_c$ of 2H-NbS$_2$ against temperature ($T$), as well as their ratio in the inset. Note that in contrast to the earlier study which reports an anisotropy of the order of 1000 (Ref. 25), our measured value was as low as 10 at room temperature, monotonically increasing to 180 on cooling. As it was shown for the case of 1T-TaS$_2$ using finite element simulations (Ref. 6), such an overestimate by the older study could be a result of a sub-optimal measurement geometry and an incorrect prior assumption that the anisotropy is very large. While $\rho_{ab}$ has a conventional metallic temperature dependence, $\rho_c$ is also metallic, but shows a few noteworthy features. First, the residual out-of-plane resistivity is very high, presumably due to a significant concentration of static defects. Second, $\rho_c$ approaches saturation in the high-temperature region. This flattening of resistivity may be attributed to the mean free paths decreasing to the point of becoming comparable to the interlayer separation, a concept known as the Mott-Ioffe-Regel limit[26]. Third, at low temperatures, $\rho_c$ displays a minimum at around 40 K, with a major upturn at lower temperatures. No corresponding feature exists in $\rho_{ab}$ (in agreement with previous results[27]). All studied samples of 2H-NbS$_2$ showed qualitatively identical behaviour, with slight differences in the absolute values of resistivity—related to slight impurity content variations—and temperatures of the minimum distributed in the 30 K – 40 K range.

We compared $\rho_c$ of 2H-NbS$_2$ to that of the isostructural and isovalent compound 2H-NbSe$_2$ (dashed line in Figure 2b). The latter material did not exhibit a similar low-temperature anomaly. Based on the nominal lattice parameters, density functional theory calculations predict that the two compounds will have nearly identical electronic band structures (see Supplementary Note 1 and Supplementary Figures 1,2). We therefore conclude that the upturn of $\rho_c$ of 2H-NbS$_2$ is not intrinsic to the nominal structure of the compound, but is caused by crystalline lattice defects.

The Seebeck coefficient ($S$) is a useful quantity for sensitively probing energy landscape variations near Fermi level. The open-circuit voltage generated by the thermal gradient is unaffected by the presence of static defects such as vacancies or non-magnetic stacking faults. On the other hand, the Seebeck coefficient is a function of the energy dependence of the conduction electron scattering rate. This is then strongly affected by the occurrence of resonance peaks in the density of states close to the chemical potential. Seebeck coefficient of 2H-NbS$_2$ revealed a prominent peak at approximately 15 K, appearing only for the out-of-plane thermal gradient (Figure 2c). The phonon drag phenomenon produces a similar feature in the temperature dependence. However, it manifests only in conductors with long phonon and charge carrier mean free paths—such as semimetals or extremely pure metals—where momentum-conserving scattering is dominant[28]. This is highly unlikely for 2H-NbS$_2$, as its high content of static defects should clearly favour momentum-relaxing scattering. Furthermore, absence of the corresponding $S_{ab}$ peak rules out the phonon drag from the possible $S_c$ anomaly origins. An alternative interpretation of the peak, the Kondo effect, will be discussed further below.

The out-of-plane resistivity anomaly of 2H-NbS$_2$ demonstrated a particularly curious response to magnetic fields. As can be seen in Figure 3a, the transverse and longitudinal out-of-plane magnetoresistances of the material are strikingly different. Applying the field along the $c$ axis suppresses the resistivity upturn, shifting the minimum to lower temperature. Yet even at 63 T the anomaly is still present. Consequently, at 50 K and below, $\rho_c$ decreases when magnetic field is increased, with signs of saturation appearing around 50 T (Figure 3b). In contrast, in transverse magnetic field, $\rho_c$ behaves as a more typical orbital magnetoresistance, common to metals. It is positive, and about three times weaker in magnitude, than the longitudinal one (Figure 3c) without significantly affecting the shape of the upturn in $\rho_c(T)$.

In order to emphasise the observed phenomenon's highly anisotropic nature, we also report the in-plane magnetotransport of 2H-NbS$_2$ up to 14 T, presented in Figure 4. The magnetoresistance is weak for all field directions, but similarly to $\rho_c$, $\rho_{ab}$ is also reduced by the field along the $c$ axis (clearly depicted in Figure 4b), which could be a trace of the same anomaly. The in-plane field-dependence of $\rho_{ab}$ (Figure 4c) is likely governed by orbital effects, like in the case of $\rho_c$.

**Discussion**

The question of the origin of the anomalies of $\rho_c$ in 2H-NbS$_2$ will now be addressed. A number of phenomena could result in a finite-temperature resistivity minimum in a metal. Resistivity upturns can be caused by electron-electron interactions in the presence of static disorder[29–31]. However, the corresponding quantum mechanical correction is either weakly enhanced by a magnetic field, or is effectively field-independent. A closely linked phenomenon of weak localisation (WL) is also known to produce an additional contribution to resistivity at low temperatures[32]. In this scenario, when a series of scattering events cause an electron to follow a closed path, quantum interference favours the net backward scattering over the forward one. Magnetic flux threading these scattering loops shifts the phases of the wavefunctions, diminishing the effect. In our case, when electrons are scattered between different planes, the closed paths should have comparable projections along the in-plane and out-of-plane directions. However, the upturn is only influenced by the $c$-axis-oriented field, contradicting the WL-based interpretation. A metal-insulator transition or conduction based on a thermally activated hopping between defects[33] would cause a divergence of $\rho_c$ at the lowest temperatures, which was not the case. The possibility of quantum tunnelling playing a significant role is ruled out based on a linear relation between current and voltage (Supplementary Note 2 and Supplementary Figure 3). Resistivity upturns have also been observed in strongly doped cuprate superconductors[34,35]. In those materials, the effect is believed to be caused by scattering from magnetic droplets forming around non-magnetic impurities. This interpretation, however, relies on the existence of strong electronic correlations, and therefore does not apply to our system.

Finite-temperature resistivity minimum in a metal is also a well-known signature of the Kondo effect; a scattering of conduction electrons off dilute localised magnetic moments[36]. Besides the upturn, the characteristic features of the phenomenon, observable in charge transport, include a negative curvature of $\rho_c(T)$ at the lowest temperatures and a suppression of the upturn by magnetic field, which causes spin-flip scattering to become inelastic[37,38]. The observed peak in the Seebeck coefficient is also characteristic to dilute and concentrated spin systems and, Kondo lattices[39,40]. It originates from the resonant scattering in the Kondo channel at the Fermi level. Above the Kondo temperature, $T_K$, the resonance is smeared out, and depending on the specifics of a system, the peak in $S$ appears at a temperature between $0.3T_K$ and $0.9T_K$ (Refs. 40, 41). When temperature is low enough, the localized spins are screened and the excitations obey simple power laws, like those of a Fermi liquid. For example, $S$ varies as $T/T_K$ for $T/T_K < 0.1 - 0.15$ for several typical Kondo alloys in the dilute, single impurity limit[42].

Measurements of $\rho_c$ under high pressure, presented in Supplementary Note 3 and Supplementary Figure 4, show that the upturn remains extremely robust up to the highest achieved pressure of 1.9 GPa. Applying pressure weakly shifts the minimum of $\rho_c$ up in temperature. This is consistent with the behaviour expected from Kondo systems[43,44].

We therefore argue that scattering off magnetic impurities is the most fitting explanation of our observations. The temperature dependence of $\rho_c$ in 2H-NbS$_2$ is consistent with the one expected from the numerical renormalisation group (NRG) theory calculations for Kondo effect[37,45], as illustrated by the fit in Figure 3a. We modelled $\rho_c$ with a sum of three contributions: a temperature-independent residual resistivity $\rho_0$, an electron-phonon scattering term $\rho_{ep}$ (captured by the Bloch-Grüneisen formula), and the Kondo term $\rho_K$ (the only magnetic-field-dependent term), for which we used the common empirical expression closely following the results of the NRG theory[37,45]:

$$\rho_K(T) = \rho_{K0}\bigl(1 + (2^{1/\alpha} - 1)(T/T_K)^2\bigr)^{-\alpha}.$$

The fitting procedure is described in more detail in the Supplementary Note 4, with the help of Supplementary Figures 5, 6, and Supplementary Table 1.

This explanation immediately raises a question regarding the nature of our system's magnetic impurities. The standard scenario where magnetic atoms are uniformly distributed clearly does not fit our picture. Doping 2H-NbS$_2$ with Fe results in the upturn observable in $\rho_{ab}$ as well as the disappearance of superconductivity[46]. Additionally, the undoped material does not display a corresponding signature in the heat capacity[47]. Lack of pronounced anomaly effects on $\rho_{ab}$ implies that the responsible defects take form of sparse planes, extending along the layers. When the current then flows along the layers, only a small fraction of the conduction electrons move in close proximity to these planes. But for the out-of-plane current flow, effectively all charge carriers have to pass through them, resulting in a particularly strong influence. Although we observed planar irregularities in the crystalline lattice via transmission electron microscopy, their atomic structure could not be determined due to a limited resolution (see Supplementary Note 5 and Supplementary Figures 7,8). Taking a closer look at the in-plane magnetoresistance for the $c$-axis-oriented field reveals that the difference between $\rho_{ab}$ at 0 and 14 T increases linearly with respect to $\ln(T)$ between 30 K and 15 K (Figure 4a inset), which further supports our hypothesis. This means that a minute contribution of Kondo scattering is present in $\rho_{ab}$, but it is not strong enough to change the sign of the gradient of $\rho_{ab}(T)$.

The presence of 1T-NbS$_2$ layers, evidenced by the characteristic CDW signatures[23], offers a fascinating interpretation of our findings, illustrated in Figure 5. As we mentioned in the introduction, the $\sqrt{13} \times \sqrt{13}$ CDW order, associated with the 1T polytype, forms a triangular superlattice of David-star-shaped clusters defined by 13 Nb atoms[24]. The electronic structure of monolayer 1T-NbS$_2$ as well as 1T-NbSe$_2$ in such a configuration has been predicted to contain one very flat band around the Fermi level. This makes the materials susceptible to electronic instabilities like Mott localisation, with a concomitant magnetic order[11,48,49]. The referenced works found the ferromagnetic insulating state as the most stable, although others have proposed that such triangular lattices can host antiferromagnetic spin-liquid phases[50,51]. These magnetic planes play the role of scatterers in the Kondo effect. The described scenario is conceptually similar to the Kondo effect occurring in artificially fabricated magnetic tunnel junctions[52,53], yet in our case the phenomenon is observed in a spontaneously formed system. The same kind of Kondo interaction has been very recently observed in a 2H/1T-NbSe$_2$ heterostructure, grown by molecular beam epitaxy[54]. One outstanding question is the anomaly's markedly different response to the two orientations of magnetic field. This difference is probably coming from the localised electron's highly anisotropic g-factor, causing a very small spin splitting (less than $T_K$), but could also be related to the magnetic ordering. Sizeable anisotropy of the g-factor is expected for systems with strong spin-orbit coupling, such as TMDs[51].

Since 2H-NbS$_2$ is known to contain frequent 3R-like stacking faults, it is natural to ask whether the anomaly is somehow caused by the inclusions of 3R-NbS$_2$. We measured the latter compound's out-of-plane resistivity, and while the corresponding temperature dependence was surprisingly found to be non-metallic, the extremely weak reaction of the interlayer conduction to the longitudinal magnetic field ($\Delta\rho_c/\rho_c \approx 0.1\%$ at 14 T) was incompatible with the behaviour observed in 2H-NbS$_2$ (see Supplementary Note 6 as well as Supplementary Figure 9 for the relevant data on 3R-NbS$_2$). The abundance of these stacking faults could explain the high residual component of $\rho_c$. The current understanding is that the poor conductivity of 3R-NbS$_2$ is not intrinsic, but rather originates from the disorder due to self-intercalated Nb atoms[55]. However, in 2H-NbS$_2$ the abundance of stacking faults results in high residual component of $\rho_c$ and good in-plane metallicity.

In summary, we have demonstrated that a delicate alternation of the interlayer crystalline structure of 2H-NbS$_2$ by introducing different polymorphs of the same atomic composition, dramatically affects the material's physical properties. In particular, the crystal's 2H stacking is occasionally disrupted by the 1T layers, which undergo a CDW instability. This then results in a triangular superlattice of David-star-shaped clusters, each hosting a lone spin at the centre. Such a texture of localised magnetic moments can be seen as a two-dimensional Kondo lattice, immersed into the metallic bulk of the 2H polytype. When the electric field or thermal gradient are then applied along the *c* axis, electronic transport shows a pronounced Kondo effect manifesting as anomalies in the out-of-plane resistivity and Seebeck coefficient. But when they are applied within the plane, there is no sign of spin dependent scattering. The observation of this highly anisotropic phenomenon occurring in a naturally formed heterostructure has been made possible thanks to the careful tailoring of the crystal by FIB. Our work therefore shows the importance of adopting new experimental techniques in studying novel electronic materials, especially highly anisotropic Van der Waals structures.

**Methods**

**Focused ion beam microfabrication.** Microstructured samples were extracted from monocrystalline flakes of TMDs. The starting crystals had the lateral size of the order of 1 mm and were at least 100 µm thick. After identifying a clean region on a crystal's surface, free of cracks or buckling, a rectangular lamella was defined by milling away the surrounding material using an FEI Helios G4 Xe plasma FIB microscope. The typical dimensions of a lamella were around 120 µm × 60 µm × 5 µm (with up to 20% variations in lengths between different samples), with the intermediate dimension corresponding to the extent along the $c$ axis of a crystal. The milling current for this stage was 60 nA, with 30 kV column voltage. An FEI Helios G3 Ga FIB microscope was then used for polishing the surface of the lamella with a 1 nA beam in order to ensure the parallelism of the two largest faces. After extraction, the lamella was glued to a sapphire substrate with a tiny amount of Araldite Rapid epoxy, keeping the external face exposed. Besides anchoring the lamella, the epoxy also formed a meniscus around it which smoothly connected the substrate's surface to the lamella's exposed face. The setup was then sputter coated with a 100 nm layer of gold. Next, the Ga FIB milling at 10 nA was used for defining the probing electrodes by selectively removing the sputtered gold layer, and for patterning the lamella in order to form the current channel and voltage probing points. The procedure was concluded with polishing the exposed side faces of the sample with a 1 nA ion beam in order to clean the surface of the re-deposited material and define the final dimensions of the device. Since the entire bottom face of the sample was rigidly attached to the substrate, differential thermal contraction and compressibility were expected to produce inhomogeneous stresses throughout the lamella. In our study, these stresses did not have a significant influence on the measured data. More detailed information about the FIB-assisted sample preparation can be found in the relevant review paper[7] and references therein.

**Resistivity measurements.** Resistivity was measured via the four-point technique with direct or alternating excitation currents in the 20–40 µA range. Temperature sweeps rate was limited to 1 K/min for the ambient pressure and of 0.5 K/min for the high-pressure measurements in order to reduce the thermal lag and gradients.

Resistivity at high pressure was measured using a piston cylinder cell produced by C&T Factory. Daphne oil 7474 was used as a pressure-transmitting medium. Pressure was determined from the changes in resistance and superconducting transition temperature of a sample of Pb located next to the 2H-$NbS_2$ sample.

Measurements in high magnetic fields were conducted at the high magnetic field facilities in Grenoble (up to 34 T DC field) and Toulouse (up to 63 T pulsed field). Quantum Design PPMS was used for measurements in fields up to 14 T.

**Seebeck coefficient measurements.** Seebeck coefficient was measured using an in-house setup. For the in-plane Seebeck coefficient measurement, a thin and long sample was mounted on a ceramic bar. One end of the bar was connected to the thermal bath, while the other one had a resistive heater attached. A differential thermocouple was used to measure the temperature difference across the sample. The out-of-plane Seebeck coefficient measurement was performed using a setup displayed in Figure 2c and described in the corresponding caption.

## Data availability

The data that support the findings of this study are available from the authors (E.M. and K.S.) upon reasonable request.

## Acknowledgements

We would like to express gratitude to Dr. Osor S. Barišić (Institute of Physics in Zagreb), Prof. Andrew Mackenzie (MPI CPfS Dresden), Prof. Fakher Assaad (University of Würzburg), Prof. Frederic Mila (EPFL), Dr. Reza Zamani (EPFL) and particularly Prof. John Cooper (University of Cambridge) for valuable discussions and feedback. We acknowledge the support of Dr. Gaetan Giriat (EPFL) concerning the instrumentation and high-pressure cells, Dr. Maja Bachmann (MPI CPfS Dresden) for her assistance with FIB microfabrication, Dr. Wen Hua (David) Bi and Davor Tolj (EPFL) for their aid with the characterisation of crystals, Dr. Diego Pasquier (EPFL) for auxiliary numerical calculations. We acknowledge the support of the European Magnetic Field Laboratory (EMFL) for access to a 34 T static magnet at LNCMI-CNRS in Grenoble (Proposal: GMA04-217), and access to a 70 T pulsed magnet at LNCMI-CNRS in Toulouse (Proposal: TSC05-119). This study has been funded by the Swiss National Science Foundation through its SINERGIA network MPBH and grants No. 200021_175836 and PP00P2_170544. C.Putzke. and P.J.W.M. acknowledge the support by the European Research Council (ERC) under the European Union's Horizon 2020 research and innovation programme (grant agreement No 715730) and the Max-Planck-Society. S.N.Z, Q.S.W. and O.V.Y. acknowledge support from NCCR Marvel.

**Author contributions**

E.M. and K.S. prepared and conducted resistivity and Seebeck coefficient measurements and wrote the manuscript together with L.F. C. Putzke, M.K., and P.M. assisted with the FIB fabrication process. H.B. synthesised the crystals used in the study. D.L. assisted with resistivity measurements at the high DC magnetic field facility in Grenoble. M.L., and C. Proust conducted resistivity measurements at the pulsed magnetic field facility in Toulouse. A.A. secured the magnet time for the experiments in Toulouse. H.K. and C.K. conducted the TEM study. S.Z., Q.W., and O.Y. provided theoretical support and the DFT data. L.F. is the project leader.

**Figures**

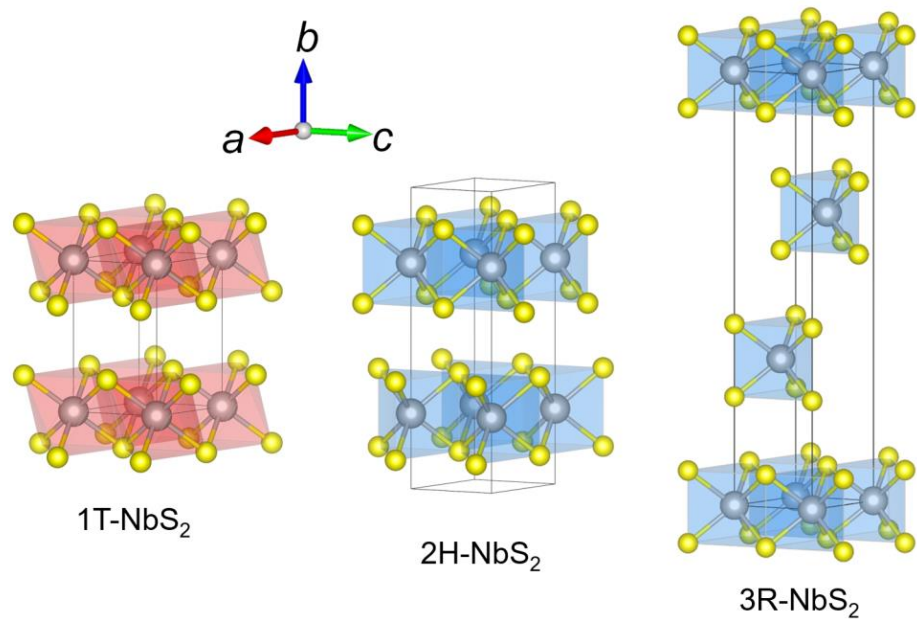

**Figure 1. Polytypes of NbS$_2$.** Crystalline lattices of 1T-, 2H- and 3R-NbS$_2$. The corresponding 1-, 2-, and 3-layer unit cells are marked with black wireframes. 2H-NbS$_2$ and 3R-NbS$_2$ share the same in-plane structure, but have different stacking of layers. For 3R-NbS$_2$, only one Nb atom with 6 nearest S atoms are shown for the middle two layers, for a clearer illustration of the stacking. Images produced with VESTA[10].

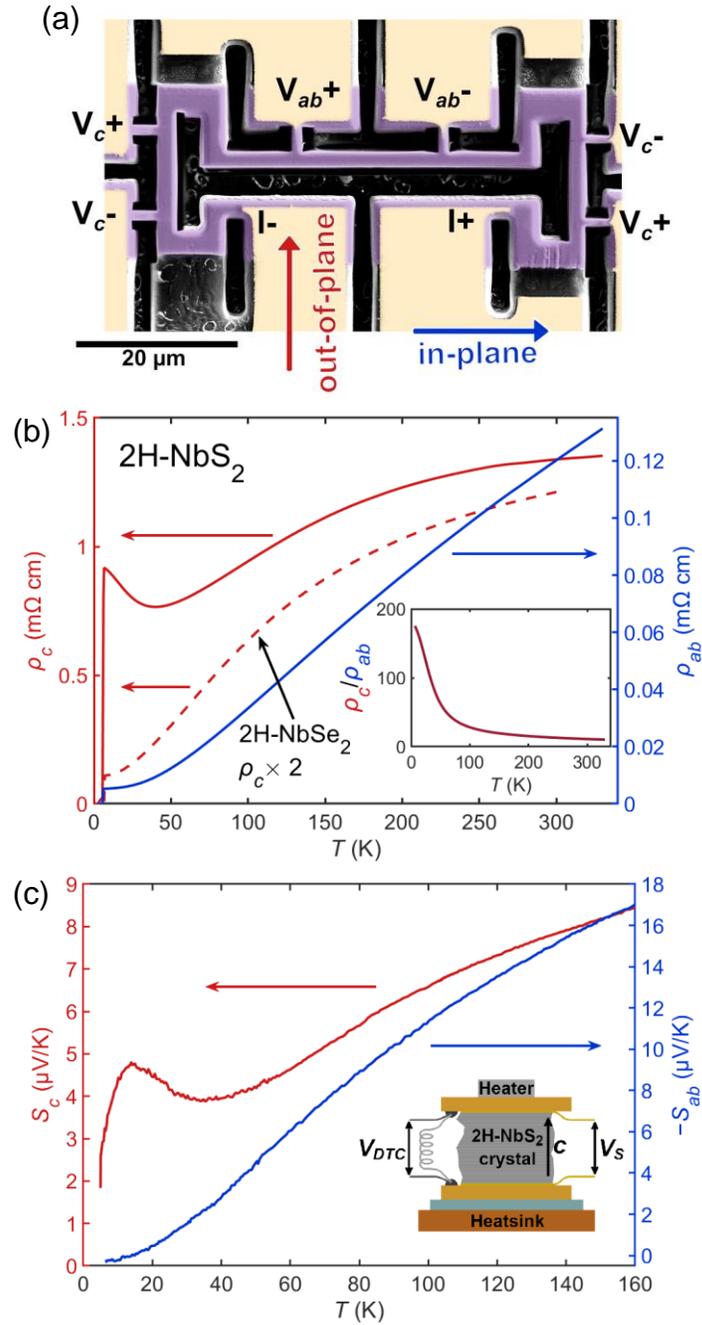

**Figure 2. Interlayer charge dynamics in 2H-NbS₂. a,** Scanning electron microscope image of a 2H-NbS₂ sample, structured with focused ion beam for accurate resistivity anisotropy measurements. False colouring is used: purple – crystal, beige – gold film. The scale bar in the top right is 20 µm long. The current sourcing (I) and voltage probing (V) electrodes are labelled. **b,** Plots of the in-plane ($\rho_{ab}$, blue) and out-of-plane ($\rho_c$, red) resistivities of 2H-NbS₂ against temperature ($T$). The dashed red line stands for the out-of-plane resistivity of 2H-NbSe₂, scaled by a factor of 2. Resistivity anisotropy of 2H-NbS₂ is plotted in the inset. **c,** Seebeck coefficients of 2H-NbS2 for the out-of-plane ($S_c$) and in-plane ($S_{ab}$) directions as functions of temperature, measured on bulk single crystals (note that $S_{ab}$ is negative). The setup for measuring $S_c$ is illustrated schematically. The crystal was approximately 1 mm long in the *c* axis direction (indicated in the drawing), and 2–3 mm long laterally. The value of $S_c$ is the ratio of the voltage across the sample ($V_s$) and the thermal gradient across it, determined from the differential thermocouple voltage ($V_{DTC}$). The sample sat between two copper plates, which homogenised temperature at its two faces and was electrically decoupled from the heatsink by a thin sapphire plate.

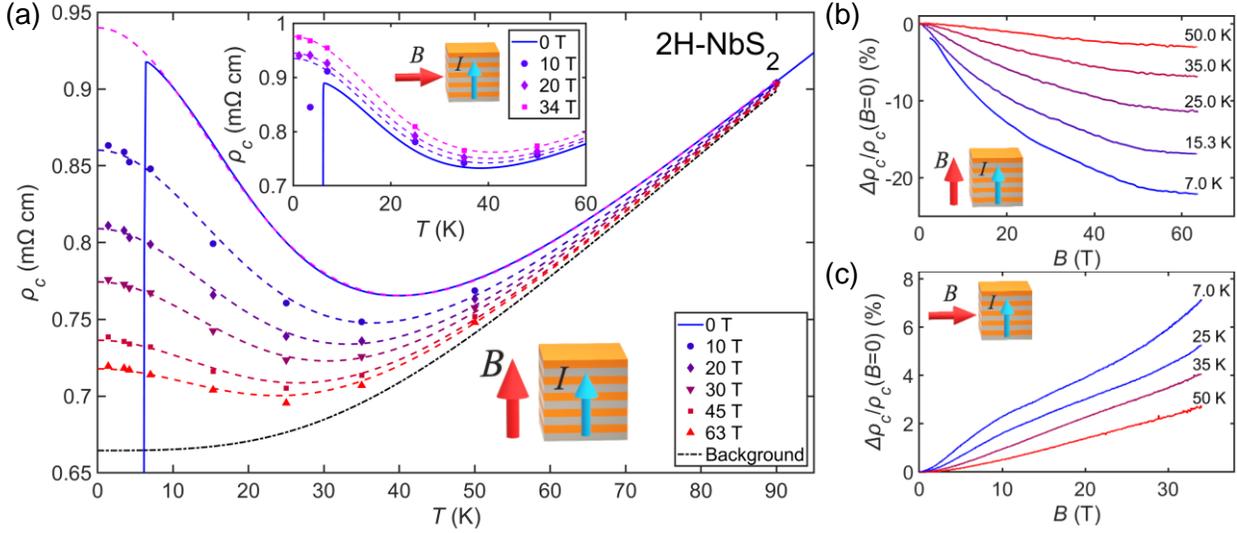

**Figure 3. Out-of-plane magnetotransport in 2H-NbS$_2$. a,** Out-of-plane resistivitiy ($\rho_c$) as a function of temperature (*T*) for various longitudinal (main plot) and transverse (inset) magnetic fields (*B*). Solid lines and markers represent the measured data. Dashed lines in the main plot are fits according to the numerical renormalization group theory of Kondo effect. The fitted model includes a field-independent contribution due to the residual temperature-independent resistance as well as the electron-phonon scattering, described by the Bloch-Grüneisen formula (dash-dot line). Dashed lines in the inset are guides for the eye. The two datasets were collected using different samples, which explains the slight resistivity mismatch for zero field. **b,c,** Relative changes in $\rho_c$, under out-of-plane (**b**) and in-plane (**c**) magnetic fields.

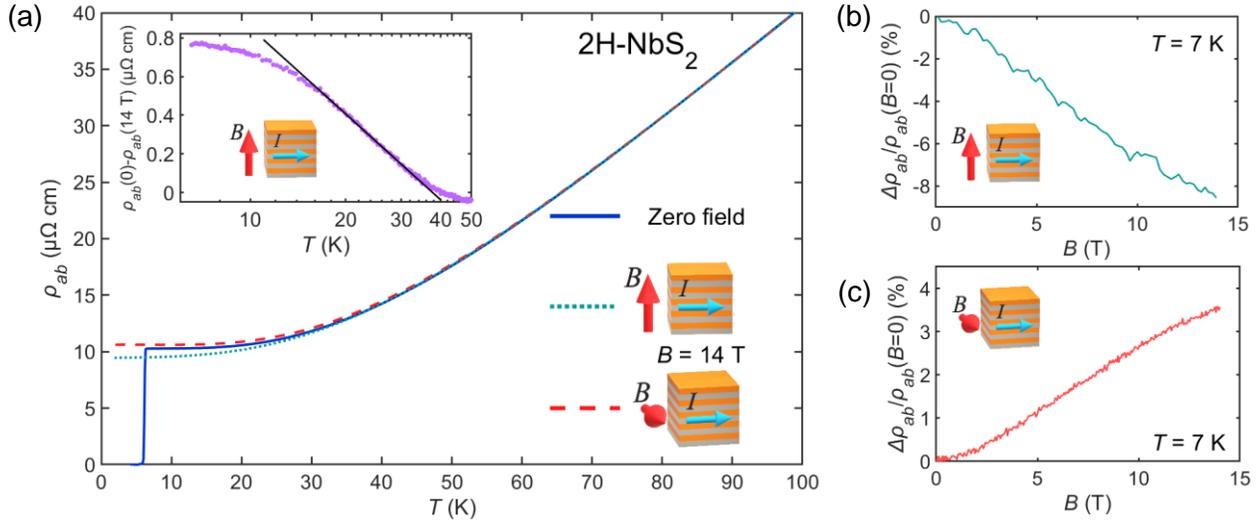

**Figure 4. In-plane magnetotransport in 2H-NbS$_2$. a,** In-plane resistivitiy ($\rho_{ab}$) as a function of temperature ($T$) at zero and 14 T magnetic field ($B$). The change of $\rho_{ab}$ between zero field and 14 T (out-of-plane field) is plotted in the inset on a logarithmic $T$ scale. The black line in the inset corresponds to the logarithmic temperature dependence. **b,c,** In-plane magnetoresistance for the out-of-plane (**b**) and in-plane (**c**) field.

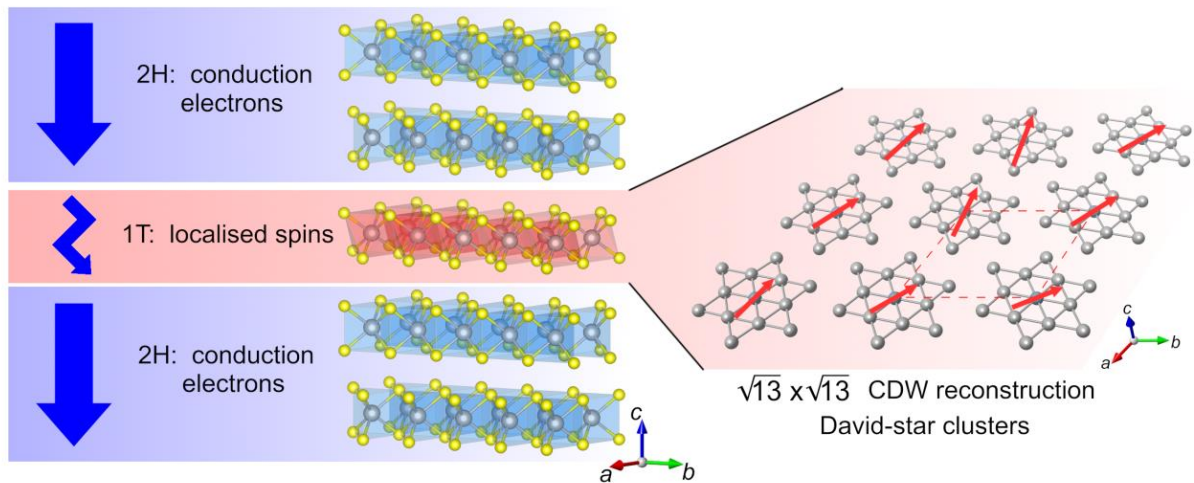

**Figure 5. 1T-NbS$_2$ inclusions in 2H-NbS$_2$.** Schematic visualisation of the proposed interpretation of the observed out-of-plane charge transport anomaly. The 2H-NbS$_2$ crystal (structure in the centre) contains inclusion layers of 1T-NbS2 (highlighted in red). 1T-NbS2 undergoes a $\sqrt{13} \times \sqrt{13}$ charge density wave (CDW) reconstruction. The Nb atoms in the CDW state are arranged into David-star-shaped clusters, superlattice of which is depicted on the right (only Nb atoms are shown, the dashed line marks the unit cell after the reconstruction). Each cluster contains an unpaired localised spin at the centre (red arrows on the right). The orientations of spins in the illustration are arbitrary and are not meant to suggest any particular ordering. This array of localised magnetic moments causes the itinerant electrons in the 2H-NbS$_2$ bulk to experience Kondo scattering during the out-of-plane current flow (blue arrows).

# Unidirectional Kondo scattering in layered NbS₂

## Supplementary Information

Edoardo Martino[1], Carsten Putzke[2], Markus König[3], Philip Moll[2], Helmuth Berger[1], David LeBoeuf[4], Maxime Leroux[4], Cyril Proust[4], Ana Akrap[5], Holm Kirmse[6], Christoph Koch[6], ShengNan Zhang[1,7], QuanSheng Wu[1,7], Oleg V. Yazyev[1,7], László Forró[1,8], Konstantin Semeniuk*[1]

1) Institute of Physics, École Polytechnique Fédérale de Lausanne (EPFL), CH-1015 Lausanne, Switzerland

2) Institute of Materials Science and Engineering, École Polytechnique Fédérale de Lausanne (EPFL), CH-1015 Lausanne, Switzerland

3) Max Planck Institute for Chemical Physics of Solids, 01187 Dresden, Germany

4) Laboratoire National des Champs Magnétiques Intenses (LNCMI-EMFL), CNRS, UGA, UPS, INSA, Grenoble/Toulouse, France

5) University of Fribourg, Department of Physics, CH-1700 Fribourg, Switzerland

6) Humboldt University of Berlin, Department of Physics, Berlin 12489, Germany

7) National Center for Computational Design and Discovery of Novel Materials MARVEL, École Polytechnique Fédérale de Lausanne (EPFL), CH-1015 Lausanne, Switzerland

8) Stavropoulos Center for Complex Quantum Matter, University of Notre Dame, Notre Dame 46556 IN, USA

## Supplementary Note 1. Electronic band structure of 2H-NbS₂ and comparison to 2H-NbSe₂

Electronic band structure and Fermi surface of perfectly ordered 2H-NbS₂ were computed by using density functional theory (DFT). The first-principles calculations have been performed using the Vienna ab initio simulation package (VASP)[1,2] within the GGA approximation. The cut-off energy of 400 eV was chosen for the plane wave basis. The lattice constants used in the calculation were $a = b = 3.31$ Å, $c = 11.89$ Å (Suppl. Ref. 3), and no additional relaxation was carried out. The band structure and Fermi surface have been obtained using the open-source code WannierTools[4] based on the Wannier tight-binding model constructed using the Wannier90 code[5].

The results are shown in Supplementary Figure 1 and are in agreement with various parts of the electronic structure of the compound presented in earlier reports[6–8]. Electronically, 2H-NbS$_2$ is nearly identical to 2H-NbSe$_2$[9,10]. Fermi surfaces of both compounds include a three-dimensional hole pocket at the Γ point of the Brillouin zone, bearing the chalcogen $p_z$ orbital character, and two kinds of quasi-tubular Fermi sheets originating from Nb $d$ orbitals. Based on these elements, particularly the hole pocket, one would expect the charge transport properties of 2H-NbS$_2$ and 2H-NbSe$_2$ to be nearly identical.

Direct comparisons between $\rho_c(T)$ and $\rho_{ab}(T)$ measured in 2H-NbS$_2$ and 2H-NbSe$_2$ are shown in Supplementary Figure 2. The values of $\rho_{ab}$ are very similar in both materials at all temperatures (Supplementary Figure 2a), which reflects their nearly identical electronic structure and comparable degrees of disorder, when individual layers are considered. Minor difference in the temperature dependences can be justified by the CDW fluctuations in 2H-NbSe$_2$ coupling to the charge transport over a broad temperature range[11]. Supplementary Figure 2b displays the stark contrast between the shapes of $\rho_c(T)$ of the two compounds. While the interval spanned by $\rho_c$ in both systems is approximately the same (in the normal states), 2H-NbS$_2$ has a much higher residual resistivity, shifting its $\rho_c$ to higher values over the entire temperature range. This implies a significantly larger content of static defects in the $c$ axis direction, compared to 2H-NbSe$_2$.

We therefore conclude that the low-temperature out-of-plane resistivity anomaly of 2H-NbS$_2$ is not an intrinsic property originating from the nominal parameters of the crystal, but rather a consequence of either frequent stacking faults (only creating disorder along the $c$ axis) or inclusions of alien phases, which extend along the layers and are rare enough to not produce signatures in the in-plane charge transport.

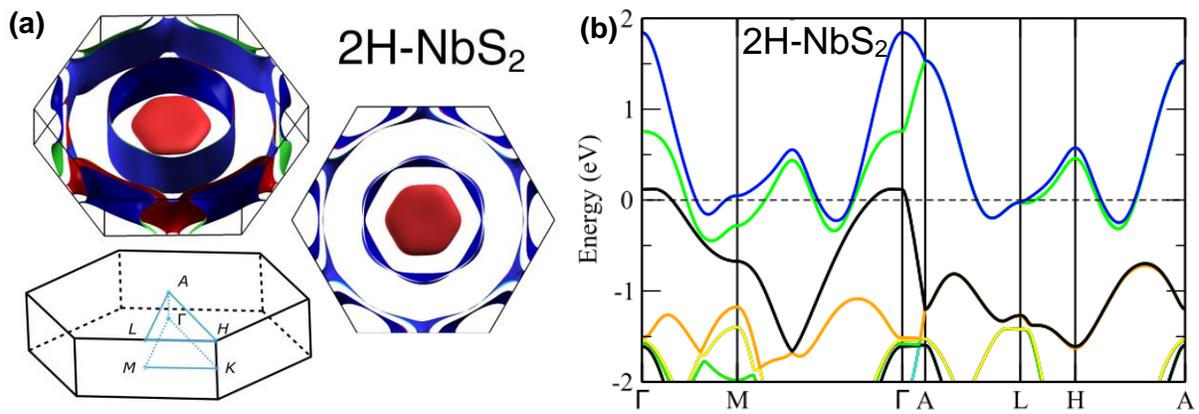

**Supplementary Figure 1. Calculated electronic band structure of 2H-NbS$_2$. a,** Fermi surface of 2H-NbS$_2$ predicted by density functional theory. Brillouin zone with high symmetry points is shown in the bottom left. **b,** The corresponding electronic band structure plots. Different bands are plotted in distinct colours (no correspondence with the colours used for the Fermi surfaces).

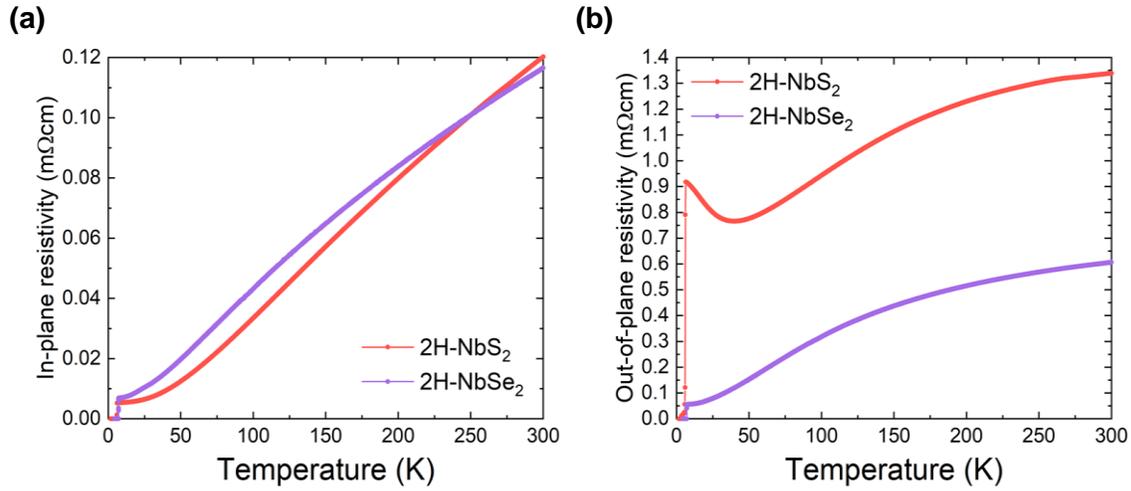

**Supplementary Figure 2. Comparison of the temperature dependences of the in-plane and out-of-plane resistivities of 2H-NbS$_2$ and 2H-NbSe$_2$. a,** In-plane resistivities for both compounds are very similar at all temperatures. In the case of 2H-NbSe$_2$, the CDW order below 35 K, and its fluctuations at higher temperatures affect the temperature dependence of resistivity. **b,** Out-of-plane resistivities, same as the ones plotted in Figure 2 of the main text, but without any multiplicative scaling. The large difference between resistivity values is evident, while both curves show a very similar temperature dependence above 50 K. The CDW transition in 2H-NbSe$_2$ has an extremely weak effect on the out-of-plane resistivity.

**Supplementary Note 2. Linearity of the out-of-plane charge transport.**

We considered a possibility that the out-of-plane resistivity anomaly in 2H-NbS2 is caused by the conduction partially occurring via a quantum tunnelling across non-conducting defects. Such a conduction results in a non-linear IV curve, however out measurements of it showed no detectible deviations from linearity (Supplementary Figure 3)

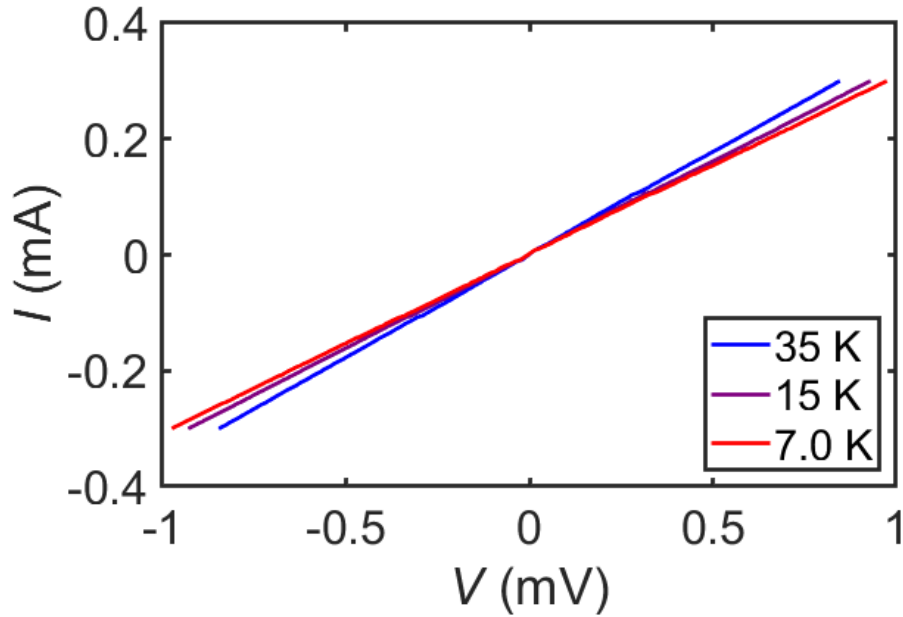

**Supplementary Figure 3. IV curve for the out-of-plane conduction in 2H-NbS$_2$.** No deviation from linearity was detected in the explored low-temperature region. Small kinks in the curves are artifacts due to the range switching of the measurement device.

**Supplementary Note 3. Pressure dependence of the out-of-plane resistivity of 2H-NbS$_2$.**

Hydrostatic pressure was another parameter used for exploring the nature of the upturn in $\rho_c$. (Supplementary Figure 4). Across the pressure range between 0.27 GPa (sealing of the pressure cell) and 1.9 GPa (the highest achieved pressure), the overall shape of $\rho_c(T)$ stayed almost the same, with $T_{min}$ shifting up by a few Kelvin. At the same time, resistivity went down with pressure at a sizeable rate of around 15% per GPa, which is presumably related to the high compressibility of the lattice along the $c$ axis (3% reduction of the lattice constant at 1.9 GPa (Suppl. Ref. 12)).

Temperature of the resistivity minimum goes up with pressure (Supplementary Figure 4 inset) like in the textbook case of Fe-doped Au[13,14]. The response to pressure is governed by a subtle balance between the decrease of the Debye temperature and the increase of the hybridization of conduction electrons with the localized spins. Usually, the latter term wins and the resistivity minimum shifts up in temperature.

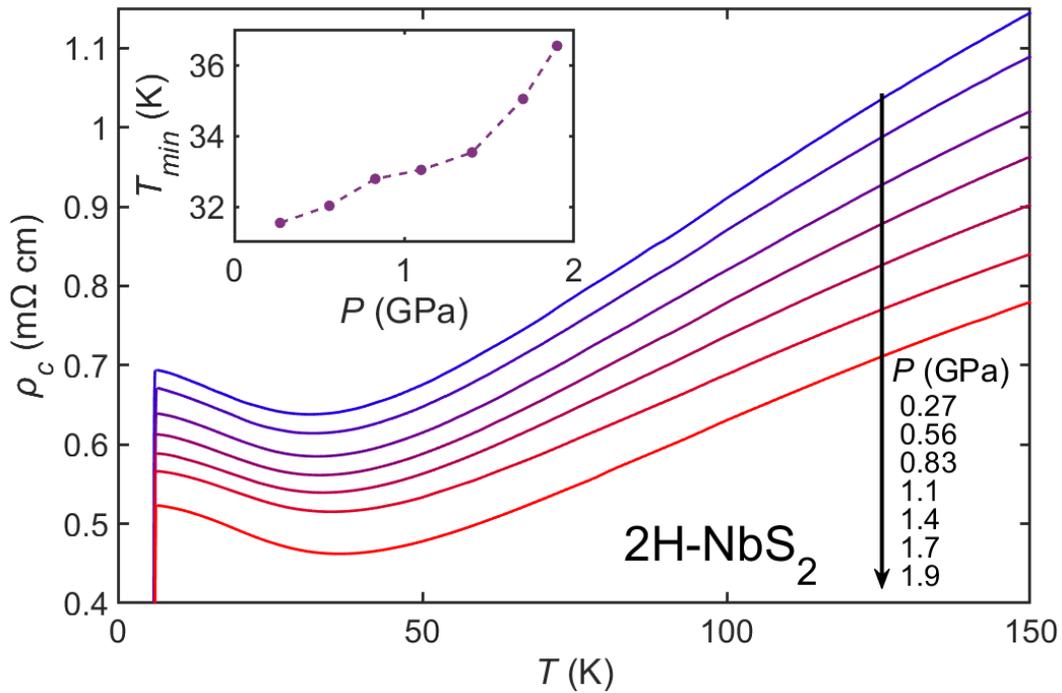

**Supplementary Figure 4. Resistivity anomaly under high pressure.** Out-of-plane resistivity of 2H-NbS$_2$ ($\rho_c$) as a function of temperature ($T$) for different values of applied hydrostatic pressure ($P$). Reduction of the lattice constants results in around 1% change of the geometrical factor, and has therefore been neglected when calculating resistivity. The estimated uncertainty in pressure is approximately 10%. The inset shows how the temperature of the minimum of $\rho_c$ shifts with pressure.

**Supplementary note 4. Fitting the out-of-plane resistivity of 2H-NbS$_2$ with a Kondo effect model.**

Here we provide a detailed explanation of the approach used for fitting the temperature dependences of the out-of-plane resistivity at different magnetic fields with the theoretical model for Kondo effect.

The overall temperature dependence of the out-of-plane resistivity is determined by multiple contributions. These can be separated into the conventional part, with a typical metallic resistivity, monotonically growing with temperature, and the anomalous component, resulting in the low-temperature upturn. The analysis is simplified if the anomalous contribution can be easily separated from the conventional one, by using the high-temperature section of the data (unaffected by the anomaly) for constraining the conventional part. In the case of 2H-NbS$_2$ this approach does not work, since the effect of the anomaly is noticeable up to about 90 K, but just above this temperature the resistivity saturation already starts coming into effect.

We therefore focused on the section of the data below 90 K and assumed that resistivity in the selected region was dominated by three terms: a temperature independent residual resistivity $\rho_0$, a contribution due to electron-phonon scattering $\rho_{ep}$, and the anomalous component $\rho_K$.

We modelled the electron-phonon scattering term with a Bloch-Grüneisen formula[15,16]:

$$\rho_{ep}(T) = C\left(T/\Theta\right)^n \int_0^{\Theta/T} \frac{t^n}{(e^t-1)(1-e^{-t})} dt.$$

The parameter $\Theta$ is the characteristic temperature, which is usually well approximated by Debye temperature. The exponent $n$ is often equal to 5, but may take different values depending on specifics of the scattering[17]. $C$ is a temperature-independent multiplier.

For capturing the anomalous contribution, $\rho_K$, presumably arising due to the Kondo effect, we used the common empirical expression, which precisely captures the temperature dependence predicted by the numerical renormalisation group theory[18,19]:

$$\rho_K(T) = \rho_{K0}\left(1 + (2^{1/\alpha} - 1)(T/T_K)^2\right)^{-\alpha}.$$

The parameter $\alpha$ is related to the spin of magnetic impurity, $\rho_{K0}$ is the zero temperature limit of $\rho_K$, and $T_K$ is the characteristic temperature (Kondo temperature), such that $\rho_K(T_K) = \rho_{K0}/2$.

The model described above has 6 parameters that have to be fitted: $\rho_0$, $C$, $n$, $\alpha$, $\rho_{K0}$ and $T_K$. Such a large number may lead to an underconstrained fit. To alleviate this problem, the least squares fit was conducted globally, for the data at different magnetic fields values (0 T, 10 T, 20 T, 30 T, 45 T, and 63 T), with only $\rho_{K0}$ being allowed to vary with field. For $\Theta$, we used the Debye temperature value of 259 K, obtained from the low-temperature heat capacity data of 2H-NbS$_2$[20].

The resultant best fit to the data is plotted in Supplementary Figure 5a. While the model does a good job at describing the measured data, if one considers the individual components $\rho_0$, $\rho_{ep}$, and $\rho_K$ separately, then the contribution ($\rho_0 + \rho_{ep}$) appears too small compared to the level expected from the visual extrapolation of the data to higher fields. If $n$ is set to the more common value of 5, the fit becomes more realistic, however a substantial discrepancy between the measured and fitted data is visible (Supplementary Figure 5b). Setting the value of $n$ to 3 (the Bloch-Wilson formula[17]) gave a noticeably better fit than for the other choices, with only a minor deviation from the measured data at 50 K for the 10 T and 20 T data (Supplementary Figure 5c, this fit is displayed in Figure 3a of the main text). The corresponding values of the fitting parameters are provided in Supplementary Table 1. The resultant value of $A$ was vanishingly small, meaning that the term proportional to $T^2$ is not needed for adequately describing $\rho_c$ of 2H-NbS$_2$

The value of 1.71 obtained for $\alpha$ is an order of magnitude larger than the typical numbers expected for spin-1/2, 1, and 3/2 magnetic impurities (0.21, 0.16, 0.146 respectively, for the case of Fe impurities noble metals[19]). On the other hand, similarly large values have been reported by studies of Kondo effect in magnetic tunnel junctions[21,22]. Our proposed idea, that the Kondo effect in 2H-NbS$_2$ is caused by planar defects, indeed bears a resemblance to the case of a magnetic tunnel junction.

The fitting was performed with an assumption that the only effect of magnetic field was the reduction of $\rho_{K0}$ and $T_K$. However, one could expect additional longitudinal magnetoresistance, associated with the conventional metallic resistivity. To estimate the magnitude of this component, we refered to 2H-NbSe$_2$. Supplementary Figure 6 shows how the out-of-plane resistivity of 2H-NbSe$_2$ is affected by the longitudinal magnetic field. Magnetoresistance is positive and increases on cooling down, reaching nearly 10% at 10 K and 14 T. When comparing magnetotransport in different systems, one should consider the dependence not simply on *B*, but rather on $\omega_c\tau$ (where $\omega_c$ is the cyclotron frequency and $\tau$ is the mean scattering time). Since at low temperature $\rho_c$ of 2H-NbSe$_2$ is around an order of magnitude lower than that of 2H-NbS2, we would expect the magnetoresistance of the latter material to vary an order of magnitude more slowly as a function of *B*. Such a contribution to magnetoresistance would not be significant compared to the observed suppression of $\rho_K$.

To summarise, the analysis presented here shows that based on the measured temperature and magnetic field dependences of the out-of-plane resistivity of 2H-NbS$_2$, Kondo effect is a plausible interpretation of the observed resistivity upturn.

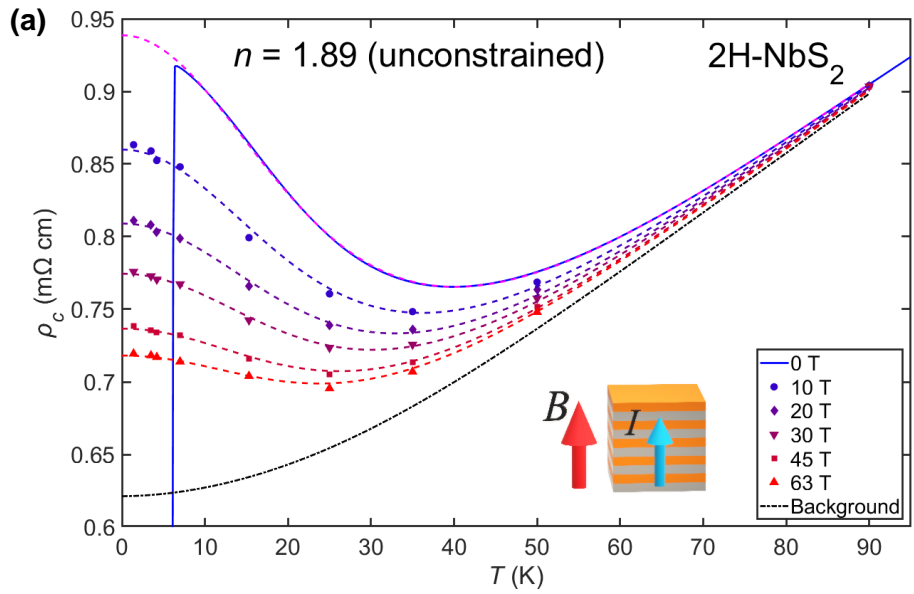
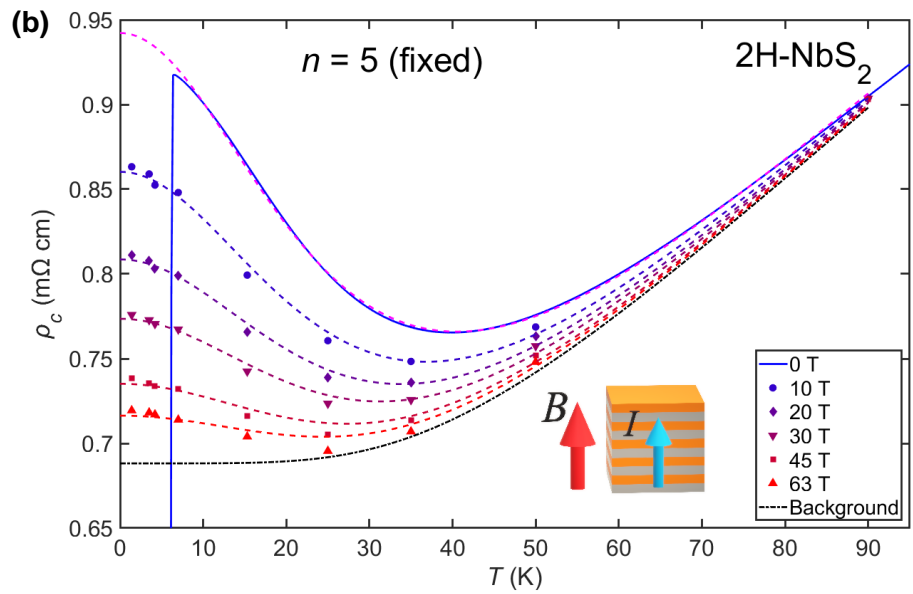
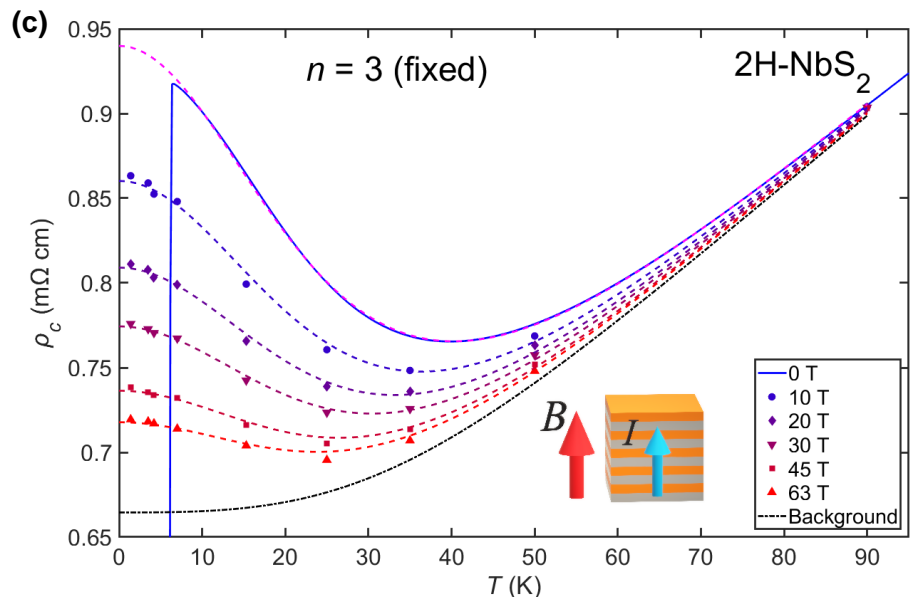

**Supplementary Figure 5. Out-of-plane resistivity of 2H-NbSe$_2$ fitted with the Kondo effect model. a–c,** Least squares fits (dashed lines) for the different values of the exponent $n$ in the Bloch-Grüneisen formula, used for capturing the conventional part of the resistivity (black dash-dot line). The solid line and the markers stand for the measured data. Leaving $n$ unconstrainted (**a**) produces a good fit, but results in an unreasonably low conventional contribution to the resistivity. Using a fixed value of $n = 5$ (**b**) causes a visible degradation of the quality of the fit. Constraining $n$ to be equal to 3 (**c**) causes a small difference in the quality of the fit, but makes the contribution due to the residual and Bloch-Grüneisen resistivities appear as a more reasonable high-field limit.

| $B$ (T) | $\rho_{K0}$ (mΩ cm) | $T_K$ (K) | $n$ | $\alpha$ | $\rho_0$ (mΩ cm) | $C$ (mΩ cm) |
|---|---|---|---|---|---|---|
| 0 | 0.317 | 23.4 | 1.88 (unconstrained) | 1.91 | 0.621 | 0.845 |
| 10 | 0.239 | | | | | |
| 20 | 0.188 | | | | | |
| 30 | 0.153 | | | | | |
| 45 | 0.115 | | | | | |
| 63 | 0.0968 | | | | | |
| 0 | 0.275 | 22.9 | 3.00 (fixed) | 1.71 | 0.665 | 1.84 |
| 10 | 0.196 | | | | | |
| 20 | 0.145 | | | | | |
| 30 | 0.110 | | | | | |
| 45 | 0.0718 | | | | | |
| 63 | 0.0532 | | | | | |

**Supplementary Table 1. Best fit parameters for the Kondo effect model applied to the out-of-plane resistivity of 2H-NbS$_2$.** Two sets of parameters are presented: with the value of $n$ obtained via fitting, and with $n$ set to 3 (giving a better agreement between the data and the model). The parameters are defined in the text.

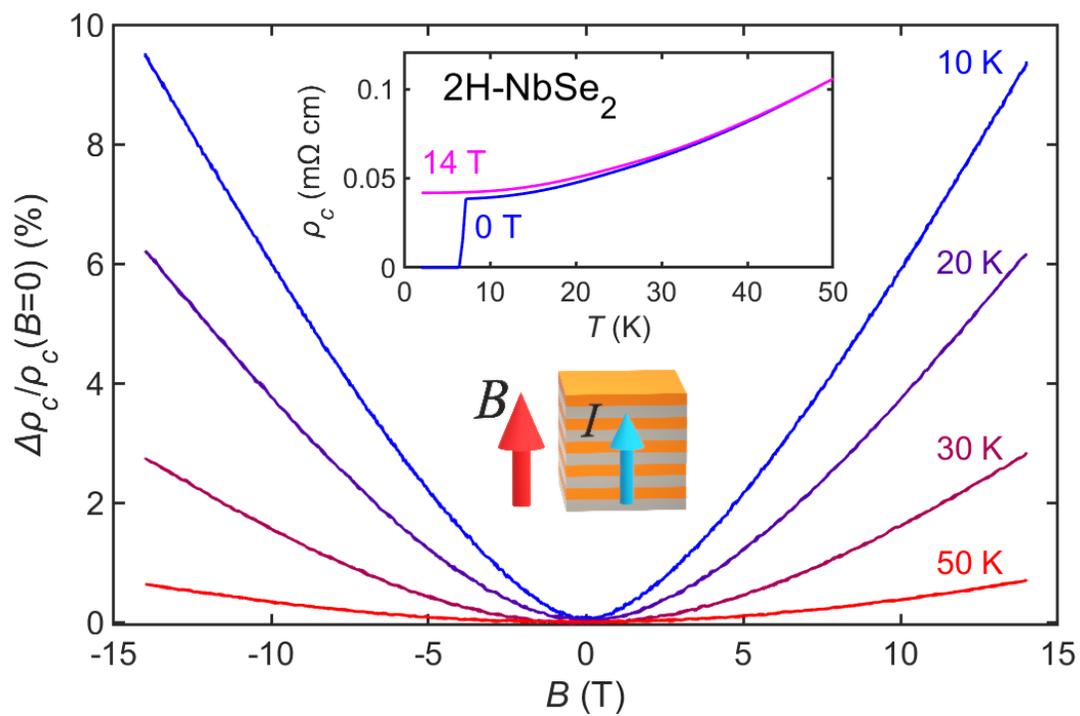

**Supplementary Figure 6. Longitudinal out-of-plane magnetoresistance of 2H-NbSe$_2$.**

**Supplementary Note 5. Evidence for the *c*-axis disorder from transmission electron microscopy.**

Transmission electron microscopy (TEM) was used in an attempt to directly detect and characterise structural defects in our crystals. The results obtained confirm their presence but allow only a limited qualitative evaluation and more work is needed to fully understand their structure and topology. For our experiments a lamella of 2H-NbS$_2$, with its plane perpendicular to a $\langle11\bar{2}0\rangle$ direction was extracted from a single crystal and polished down to sub 50 nm thickness with focused ion beam[23]. First, by analysing a transmission electron diffraction image, we concluded that majority of reflections belonged to both 2H and 3R-types of stacking, but some spots could not be accounted for (Supplementary Figure 7).

This result is in agreement with the earlier descriptions of this material[24,25]. In the 2H and 3R structures the layers are identical, but their stacking sequences are different. Due to the lower symmetry of 3R-NbS$_2$, two different orientations of it needed to be considered (green and purple dashed lines). The pattern had to be vertically stretched by about 9% in order to get the best match with the expected spot positions.

High-angle annular dark-field imaging (HAADF) scanning TEM images revealed extended homogeneous regions separated by two kinds of planar defects, comprising a single or three dark lines running along the layers (Supplementary Figure 8a). In the former case, the intensity profile on two sides of the defect was shifted by half a period, while no such change occurred for the latter one (Supplementary Figure 8b). The phase shift indicates that the single-line defect represents a layer of increased thickness. It was not possible to achieve atomic resolution and deduce the metal-chalcogenide coordination. This would have been necessary in order to identify the 1T-type layers, where metal is octahedrally coordinated by suphur, as opposed to the corner-sharing trigonal prismatic coordination for 2H and 3R polytypes.

Dark appearance of the defects could be caused by a variety of mechanisms, such as stoichiometry variation, diffraction contrast or even local magnetic order. The fact that these defects take form of isolated and laterally extended planes makes Nb deficiency an unlikely explanation, as the vacancies are expected to be distributed more randomly. A more fitting explanation would be the atomic disorder/displacement, which can originate from a 1T-NbS$_2$ layer with the $\sqrt{13} \times \sqrt{13}$ CDW order. Traces of this particular lattice reconstruction have been observed in 2H-NbS$_2$ using diffusive X-ray diffraction[25]. Increased thickness of the defective layer also agrees with the scenario proposed above, as the associated distortion of the lattice is known to increase the interlayer spacing[26]. The role of local magnetism in generating the contrast cannot be excluded. This hypothesis could be tested by further studies by TEM with differential phase contrast.

The larger than expected interlayer separation observed in the studied lamella remains an open question, and additional studies are necessary in order to make conclusions about the topology of stacking faults.

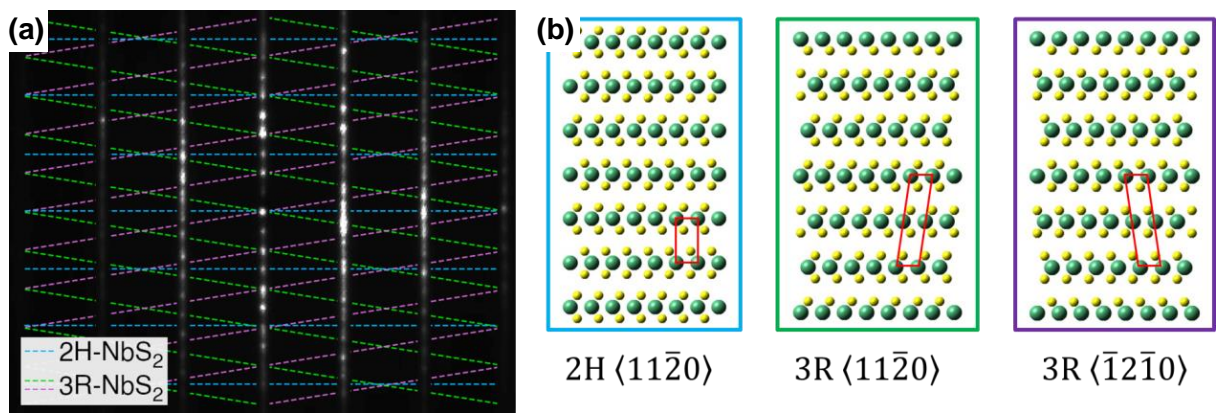

**Supplementary Figure 7. Transmission electron diffraction study of 2H-NbS$_2$. a,** Transmission electron diffraction pattern, when viewed along a $\langle 11\bar{2}0 \rangle$ direction. Dashed lines mark the rows of spots expected from different polytypes (3R-NbS$_2$ has no inversion symmetry, so two different orientations have to be considered). **b,** Details of the crystalline structure of 2H- and 3R-NbS$_2$ used to model the observed diffraction pattern. Nb and S atoms are shown in green yellow spheres, respectively.

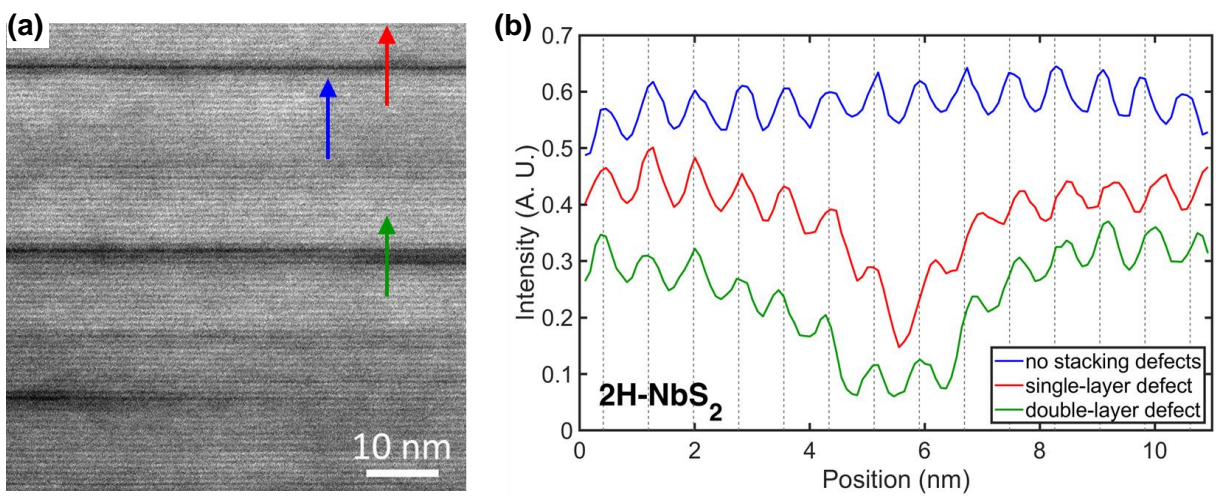

**Supplementary Figure 8. Transmission electron microscopy study of 2H-NbS$_2$. a,** High-angle annular dark field (HAADF) scanning transmission electron microscopy (STEM) image of the 2H-NbS$_2$ lamella. **b,** Intensity profiles extracted from the regions of the STEM image marked with the arrows of the corresponding colour in panel **a**.

**Supplementary note 6. Out-of-plane resistivity of 3R-NbS₂**

Since 2H-NbS$_2$ is known to contain frequent stacking faults, locally appearing as 3R-NbS$_2$, we also probed resistivity of the latter material along the two directions. The available single crystals of the 3R polytype had 1.5% of S substituted by isovalent Se. An isovalent substitution in TMDs can strongly tune the stability of often competing CDW and superconducting states, but generally has little effect on the normal state charge transport properties. Since neither 3R nor 2H-NbS$_2$ develop any CDW order, the aforementioned doping is of minor relevance. The measured out-of-plane and in-plane resistivities, as well as their ratio, are plotted against temperature in Supplementary Figure 9a. The shape of the in-plane resistivity is identical to the ones reported for undoped 3R-NbS$_2$[27–29] Low residual in-plane resistivity ratio of 1.3 implies a high content of static defects, yet similarly low values have been found in the earlier studies. Below 30 K, the in-plane resistivity exhibits a very weak upturn, commonly observed in this compound and explained by electron-electron interactions in presence of static disorder[27]. Surprisingly, the out-of-plane resistivity has a negative and nearly constant gradient throughout the entire explored temperature range. Resistivity anisotropy of 3R-NbS$_2$ goes from 6 to 9 upon cooling. The longitudinal out-of-plane magnetoresistance (Supplementary Figure 9b) is positive and extremely weak, unlike in 2H-NbS$_2$. Based on this observation, we conclude that the out-of-plane resistivity anomaly of 2H-NbS$_2$ cannot be explained in terms of inclusions of 3R-NbS$_2$.

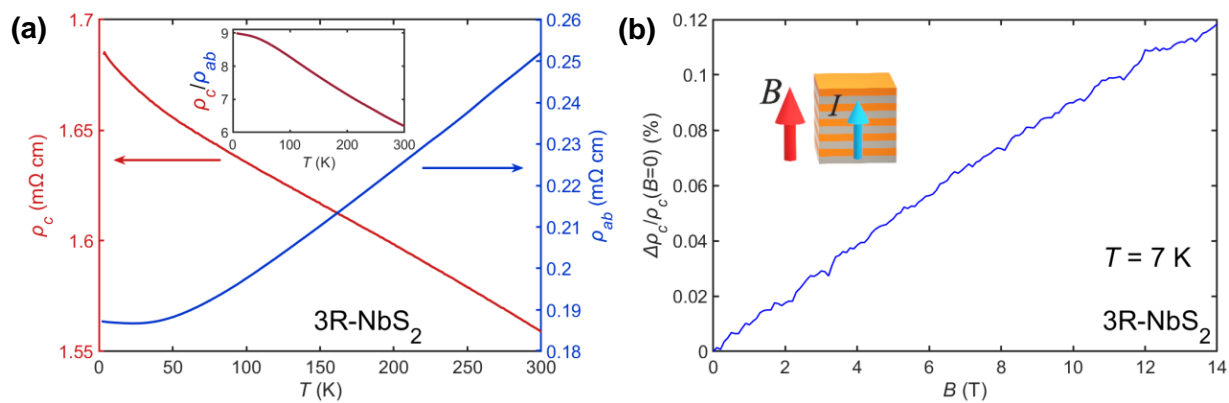

**Supplementary Figure 9. Resistivity anisotropy and magnetotransport of 3R-NbS$_2$. a,** Out-of-plane and in-plane resistivities of 3R-NbS$_2$ as functions of temperature, with their ratio plotted in the inset. **b,** Longitudinal out-of-plane magnetoresistance at 7 K.

**Supplementary References**